\documentclass{article}
\usepackage{graphicx}
\usepackage{url}
\usepackage{authblk}
\usepackage{natbib}
\title{Latitudinal Profiles of the Photospheric Magnetic Field
for Solar Cycles 19 - 21}
\author[1]{E.S.~Vernova}
\author[1]{M.I.~Tyasto}
\author[2]{D.G.~Baranov}

\affil[1]{IZMIRAN, SPb. Filial, St.~Petersburg, Russian Federation

email: helena@ev13934.spb.edu}

\affil[2]{A.F.~Ioffe Physical-Technical Institute, St.~Petersburg,
Russian Federation}

\date{}                     
\begin{document}
  \maketitle

\sloppy
\begin{abstract}
For groups of photospheric magnetic fields differing in strength
the time averaged synoptic maps were obtained on the basis of the
National Solar Observatory Kitt Peak data (1976--2003). The
latitudinal   profiles of  magnetic field fluxes were considered
separately for each 5 G interval of the field strength B. The
changes both of the  profile maxima and of the latitudinal
localization of these maxima were studied. The obtained results
show that the latitudinal distribution of magnetic fields
considerably changes at certain values of the field strength:
$5$\,G, $15$\,G, and $50$\,G. The magnetic flux for field groups
differing in strength, monotonically decreases with increase of
strength for all fields with strength $B>5$\,G. For these fields
the fluxes of the southern hemisphere exceed  fluxes of the
northern hemisphere. A very special group is represented by the
weakest fields $B<5$\,G, which are in  antiphase both in
localization and in time changes with stronger fields.
\end{abstract}

\section{Introduction}

The interest to the latitudinal distribution of solar activity is
connected with such important problems as displacement of  sunspot
zones during 11-year solar cycle, periodic appearance of polar
faculae in high latitudes, development of polar coronal holes,
etc. (see review and references in \citep{hath}. The most studied
feature among these phenomena is the latitudinal distribution of
sunspots and its change during 11-year solar cycle. According to
Sp\"orer's law the mean latitude of sunspot groups gradually
decreases from the beginning to the end of the solar activity
cycle, i.e. sunspot zone moves from mean heliographic latitudes
towards equator of the Sun. It was shown by \cite{li}, that the
latitude migration-velocity of sunspot groups is maximal at the
beginning of a solar cycle, decreasing later in process of solar
cycle development. The width of the sunspot zone and its
connection to the solar cycle was studied in \citep{ivan, mile}.

Another manifestation of solar activity concentrating around
certain  interval of heliolatitudes, are  polar faculae
\citep[see, for example,][]{shee}. Polar faculae appear at higher
latitudes than  sunspots and precede sunspots in their time
development approximately by 6 years \citep{maka}. Another
research \citep{deng} has shown that polar faculae lead sunspot
number by 52 months. Polar faculae measurements display good
correlation with polar magnetic fields \citep{muno}.

All the wealth and diversity of the forms of solar activity
phenomena are connected to the magnetic fields of the Sun
differing both in strength and  in localization on the Sun's
surface. The evolution of zonal distribution of the solar magnetic
field was studied on the basis of Wilcox Solar Observatory (WSO)
magnetograms \citep{hoek}. It was shown, that the total magnetic
flux is intimately connected to a level of activity and its
distribution is similar to the Maunder butterfly diagram,
presenting latitude of sunspot occurrence versus time. The
variations of magnetic fields in time and latitude were considered
in \citep{akht}. The strong difference of variations of a total
magnetic flux at low and high latitudes is revealed. The analysis
of development of latitudinal distributions for magnetic fields
with different strengths in solar cycles 12--23 has shown, that
the width of a sunspot zone is intimately connected with the total
magnetic flux of sunspots \citep{mile}.

In our paper \citep{vern} the latitudinal distribution of magnetic
fields was studied on the basis of synoptic maps of the
photospheric magnetic field obtained by the Kitt Peak National
Observatory (1976--2003). To consider the latitudinal distribution
of magnetic fields we used the method of synoptic map
superposition. As a result one average synoptic map of the
magnetic field module for all period of 1976--2003 was
constructed. Averaging magnetic field over the longitude, we
received a latitudinal  profile of magnetic flux of the Sun for
three solar cycles. Thus, the stable features of the latitudinal
distribution of magnetic field appear which are present during
several solar cycles. In the averaged latitudinal profile two
areas of magnetic field concentration were detected: one at the
sunspot zone latitudes and the other at the polar faculae zone.
The flux in the sunspot zone considerably exceeds that in the
polar faculae zone. It is necessary to note, that the latitudinal
profile in all range of latitudes has some lower threshold of the
flux $~1.3\times10^{21} Mx$. The maximum flux in the sunspot zone
exceeds this threshold value approximately by 5 times. The results
obtained by us display approximately symmetrical latitudinal
distribution of the magnetic field module relative to
helioequator. However the flux of the southern hemisphere,
averaged for three solar cycles, proved to be somewhat higher than
the flux of the northern hemisphere.

The main purpose of this research was to observe features of a
latitudinal distribution of magnetic fluxes for different values
of the magnetic field strength. Special attention was given to the
distribution of weak magnetic fields ($B<50$\,G). For these field
strengths the most significant differences in latitudinal profiles
of photospheric magnetic fields are observed.

\section{Data and method}

In this study synoptic maps of the photospheric magnetic field
produced by the National Solar Observatory (NSO Kitt Peak) for
1976--2003  were used (http://nsokp.nso.edu/). Synoptic maps were
obtained with the resolution of $1^\circ$ in longitude (360 steps)
and 180 equal steps in the sine of the latitude. Thus, each map
contained $360\times 180$ pixels of  magnetic field strength in
Gauss.

In order to underline the steady features of a latitudinal
distribution, synoptic maps were averaged for three solar cycles.
For this purpose we  used the method of synoptic map
superposition, which has allowed to receive one average synoptic
map for all period of 1976--2003.  Only absolute values of the
magnetic fields were considered without taking into account their
polarity.

To estimate the contribution of fields of different strength to
the total magnetic flux, each synoptic map before summing was
transformed in such a manner that only pixels in a chosen interval
of field strength were left, while the remaining pixels were
replaced by zeros. Thus, we obtain a summary map for three solar
cycles for each particular interval of magnetic field strength.
The number of pixels distinct from zero, displayed the ratio of
particular group of fields in the total flux of the magnetic
field.

Synoptic maps for different intervals of a field strength were
constructed separately: from $0$\,G  up to $5$\,G, from $5$\,G up
to $10$\,G, etc. By averaging these maps along the heliolongitude
we obtained a latitudinal  profile of  magnetic field in a
selected strength interval.

\section{Results and discussion}

\begin{figure}
\begin{center}
\includegraphics[width=0.9\textwidth]{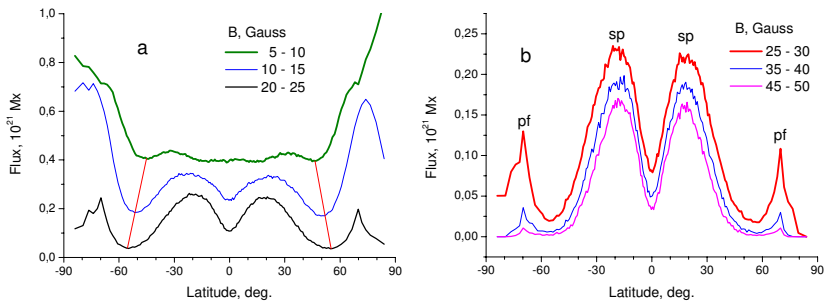}
\caption{ Latitudinal profiles of the photospheric magnetic
fields, averaged for three solar cycles (1976--2003). (a) Magnetic
fields with strength from $5$\,G to $25$\,G (straight red lines
show displacement of the profile minimum depending on the field
strength). (b) Magnetic fields with strength from $25$\,G to
$50$\,G. Regions of magnetic flux concentration are marked -
sunspot zones (sp) and polar faculae zones (pf).} \label{vernova1}
\end{center}
\end{figure}

In Figure~\ref{vernova1}a latitudinal  profiles of the magnetic
flux, averaged for solar cycles 21--23, are presented for three
intervals of the field strength: $5-10$\,G, $10-15$\,G, and
$20-25$\,G. The latitudinal profile for fields of $5-10$\,G
differs significantly in its form from the profiles of  stronger
fields (Figure~\ref{vernova1}a): the flux is almost constant from
$-60^\circ$ up to $+60^\circ$ and does not fall anywhere below
$0.39\times10^{21} Mx$. At high latitudes the magnetic flux
increases more than twice as compared with the lowest limit. In
spite of the approximate constancy of the magnetic flux at middle
latitudes, some small increase of flux at the sunspot zone
latitudes can be seen already in the latitudinal profile. With
further increase of strength ($B>10$\,G) the latitudinal profile
starts gradually to change its form. At low latitudes the
concentration of magnetic fields appears more distinctly in the
sunspot zone. At high latitudes the maximum of the flux moves from
poles to the polar faculae zone. While for the group of $10-15$\,G
the flux at high latitudes considerably exceeds the flux in the
sunspot zone, with the increase of the field strength up to
$20-25$\,G the  fluxes in the sunspot zone and in the polar
faculae zone became nearly equal (Figure~\ref{vernova1}a).
Other feature of latitudinal profiles of weak fields is the
presence of several regions of minimum field concentration, which
appear more and more clearly with increase of the field strength.
One of the flux minima is observed at near equatorial latitudes,
others two - at latitudes $40^\circ - 60^\circ$ in each of
hemispheres. The position of the equatorial minimum is constant
for all groups of fields, while in the range of latitudes
$40^\circ - 60^\circ$ the displacement of the minima occurs with
the increase of the field strength (straight red lines in
Figure~\ref{vernova1}a).

For groups of fields $25-30$\,G, $35-40$\,G and $45-50$\,G
(Figure~\ref{vernova1}b) all latitudinal   profiles have nearly
the same form with two maxima in each hemisphere: in the sunspot
zone (sp) and in the polar faculae zone (pf) at latitude about
$70^\circ$. The maximum of the flux in the sunspot zone
considerably exceeds the flux in the polar faculae zone. For the
strength interval $45-50$\,G the flux in the polar faculae zone is
practically equal to zero. Thus, the value of the field strength
$50$\,G can be considered as the boundary above which latitudinal
profile displays only the sunspot zone. It should be noted, that
in both zones with the increase of field strength the maximum of
the flux will decrease monotonically.
\begin{figure}[t]
\begin{center}
\includegraphics[width=0.9\textwidth]{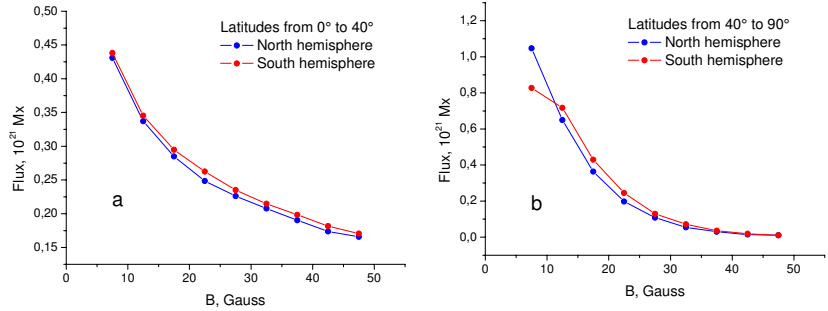}
\caption{ Change of the latitudinal profile maximum for northern
and southern hemispheres depending on magnetic field strength.
There are two maxima in the latitudinal profile for each of the
hemispheres: the values of flux maxima are presented separately
for latitudes $0^\circ - 40^\circ$ (a) and $40^\circ - 90^\circ$
(b).} \label{vernova2}
\end{center}
\end{figure}
\begin{figure}[h]
\begin{center}
\includegraphics[width=0.9\textwidth]{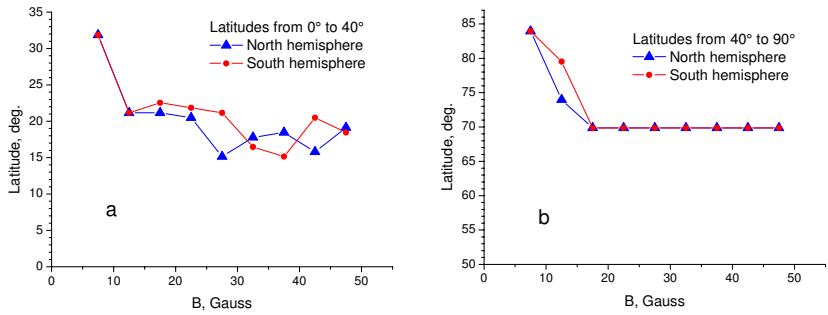}
\caption{ Regions of the highest concentration of magnetic fields
with different strength: dependence of position of the latitudinal
profile maximum on the magnetic field strength for latitudes
$0^\circ - 40^\circ$ (a) and $40^\circ - 90^\circ$ (b).}
\label{vernova3}
\end{center}
\end{figure}
In  Figure~\ref{vernova2} the relation of the flux maximum on the
magnetic field strength is shown for northern and southern
hemispheres. We consider separately: (a) low latitudes from
$0^\circ$ up to $40^\circ$ and (b) high latitudes from $40^\circ$
up to a Sun's pole.

Common features of the low latitude fields
(Figure~\ref{vernova2}a) and of the high latitude fields
(Figure~\ref{vernova2}b): the continuous decay of the flux maximum
with increase of the magnetic field strength is seen in both
cases, the flux of the southern hemisphere exceeding slightly the
flux of the northern hemisphere. With the magnetic field strength
increase the speed of the flux falling is slowed down. In
comparison with fields of $5-10$\,G the magnetic flux for fields
of $45-50$\,G decreases by 3 times (Figure~\ref{vernova2}a).

Difference between  fluxes at low (Figure~\ref{vernova2}a) and at
high latitudes (Figure~\ref{vernova2}b): the largest value of the
high latitude flux, which corresponds to the group of fields
$5-10$\,G, is twice that of the low latitudes. While for the
strongest fields at low latitudes (group of $45-50$\,G) the flux
falls only down to a level of $0.17\times10^{21} Mx$, for high
latitudes flux value drops almost down to zero with the increase
of the field strength.

Latitude of localization of the profile maximum depending on
magnetic field strength is shown in Figure~\ref{vernova3} for
northern and southern hemispheres. For low latitudes
(Figure~\ref{vernova3}a) the position  of maxima of two
hemispheres coincide for the weak fields of $5-10$\,G and
$10-15$\,G. Then northern and southern hemispheres develop
differently, but the latitudes change insignificantly and vary
between  $15^\circ$ and $22^\circ$. Thus, the maxima of the flux
for these fields are  in sunspot zones. It can be seen for high
latitudes (Figure~\ref{vernova3}b), that the weak fields $5-10$\,G
and $10-15$\,G are localized at the highest latitudes and,
apparently, are connected with coronal holes. Beginning from
strengths of $15-20$\,G both hemispheres display one dominating
latitude around $70^\circ$ which is connected to localization of
polar faculae at this latitude.

\begin{figure}[t]
\begin{center}
\includegraphics[width=0.9\textwidth]{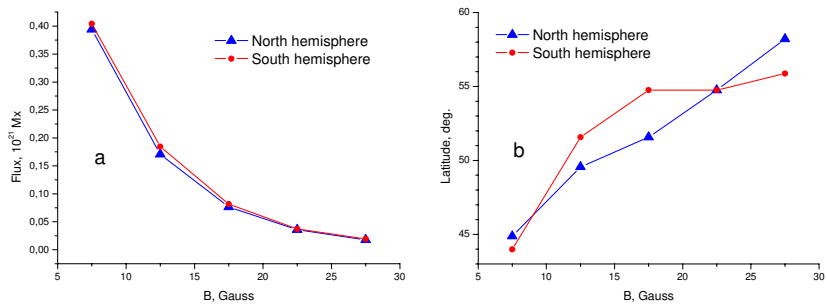}
\caption{ Regions of the lowest concentration of magnetic fields
with different strength: (a) - flux value in a minimum of the
latitudinal profile; (b) - latitudinal position of minimum.}
\label{vernova4}
\end{center}
\end{figure}

It is of interest to consider  latitude regions, where the
magnetic flux is minimal. The change of value and localization of
the flux minimum is shown in Figure~\ref{vernova4} depending on
the field strength. The minimum of a latitudinal profile
(Figure~\ref{vernova4}a) drops down with increase of the field
strength in the same way as the value of a maximum
(Figure~\ref{vernova2}a,b). The monotonic decrease of the flux
with increase of the magnetic field strength is seen. For the
whole strength range the flux of the southern hemisphere exceeds
the flux of the northern hemisphere slightly. The displacement of
latitudinal profile minimum in latitude range of $40^\circ -
60^\circ$ (Figure~\ref{vernova4}b) is observed. The flux minimum
for fields of $5-10$\,G appears at the latitude $\sim45^\circ$. As
the field strength increases, the minimum  is displaced to higher
latitudes and for fields of $25-30$\,G the minimum occurs at
latitudes $56^\circ$ (southern hemisphere) and $58^\circ$
(northern hemisphere).

It is noteworthy, that latitudinal   profiles for fields of
$5-10$\,G, $10-15$\,G etc. form the set of similar curves, whose
shapes gradually change in the course of the field strength
increase, so that the curves do not intersect anywhere
(Figure~\ref{vernova1}a,b). The weakest fields with strength
$B<5$\,G form a very special group, which is at the same time and
the most numerous group. Pixels with a field strength of $0-5$\,G
for period 1976--2003 occupy more than 60\,\% of the solar
surface, while strong fields ($B>50$\,G) occupy only 3.3\,\% of
the surface.
\begin{figure}[t]
\begin{center}
\includegraphics[width=0.9\textwidth]{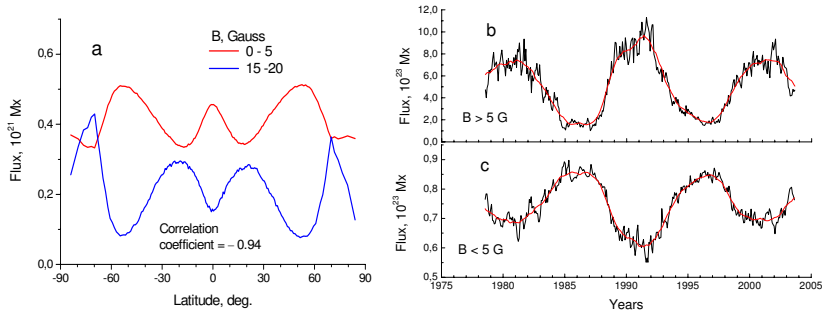}
\caption{ Comparison of the weakest fields features with those of
the stronger fields: (a) latitudinal profiles of fields of
$0-5$\,G and $15-20$\,G (correlation coefficient $R = - 0.94$);
time changes of magnetic fields $B>5$\,G (b) and $B<5$\,G (c)
(correlation coefficient $R = -0.96$). The red line represents
moving averages over 20 Carrington rotations.}
\label{vernova5}
\end{center}
\end{figure}

While studying the weakest magnetic fields ($B<5$\,G) the
influence of the instrumental noise  becomes essential. The
high-latitude measurements are less reliable because the magnetic
field is radial in the photosphere and has only a small component
in the line-of-sight direction where the instrument is sensitive.
According to \citep{harv}, the instrumental noise level in the
polar regions in NSO/KP maps is a function of latitude and is on
the order of $2$\,G per map element. The influence of these
measurement errors is diminished by the data treatment we used
here. To obtain latitudinal profiles, synoptic maps were averaged
in longitude and then averaged as a function of time for nearly
three cycles. With this averaging the noise level is reduced
significantly.

In Figure~\ref{vernova5}a a comparison of the latitudinal profile
of the  weakest fields ($B<5$\,G) with a  profile of stronger
fields is carried out (as an example the latitudinal profile of
$15-20$\,G  fields is chosen). For the latitudinal profile of
$15-20$\,G, flux concentration  in sunspot and polar faculae zones
is observed, which is characteristic for all groups of fields with
strength above $15$\,G. The magnetic fields with strength $B<5$\,G
develop in antiphase and display minima of the flux both at
latitudes of the sunspot zones and at latitudes of the polar
faculae zones. These fields concentrate at latitudes from
$40^\circ$ up to $60^\circ$ and in equatorial region $\pm5^\circ$.
The maxima of the weakest fields are located at latitudes
$53^\circ$ and at the equator. Thus, two latitudinal profiles
($15-20$\,G) and ($B<5$\,G) are the reverse of each other, with
correlation coefficient of two curves $R = - 0.94$.

Not only the latitudinal profile for the weakest fields ($B<5$\,G)
is in antiphase to the profile of stronger fields, but also time
course of these fields develops in antiphase with strong fields.
In Figure~\ref{vernova5}b the time course of magnetic flux for
fields with strength $B>5$\,G is shown. The pronounced 11-year
cycle is seen. The maximum of the flux coincides with the second
Gnevyshev maximum.

In Figure~\ref{vernova5}c the time course for fields of $B<5$\,G
is shown. It can be seen, that  weak fields develop in antiphase
with stronger fields (coefficient of correlation $R = -0.96$),
with the maximum values of  flux approximately by the order lower,
than for strong fields. This result is in agreement with results
obtained on the basis of SOHO/MDI data for 1996--2011, which show
that magnetic structures with a weak flux display anticorrelation
with sunspots \citep{jin}.

\section{Conclusions}

The form of the latitudinal profile  changes as the magnetic field
strength increases and these changes follow some  regularity. The
absolute flux value decreases both at maximum and at minimum of
the profile along with the increase of  the field strength. The
localization of the profile extrema also depends on the field
strength. While the latitude of the profile maximum does not
change significantly, the minimum of the flux monotonically moves
towards higher latitudes with increase of the magnetic field
strength.

The obtained results show that the latitudinal distribution of
magnetic fields considerably changes at certain values of the
field strength: $5$\,G, $15$\,G, and $50$\,G. The magnetic flux
for field groups differing in strength, monotonically decreases
with increase of strength, while the fluxes of the southern
hemisphere exceed  fluxes of the northern hemisphere. A very
special group is represented by the weakest fields $B<5$\,G, which
are in antiphase both in localization and in time changes with
stronger fields.

\section{Acknowledgements}

NSO/Kitt Peak magnetic data used here are produced cooperatively
by NSF/NOAO, NASA/GSFC and NOAA/SEL.

\end{document}